\documentclass[12pt,english]{iopart}
\usepackage{graphicx}
\usepackage{amssymb}

\makeatletter
\usepackage{iopams}
\usepackage{setstack}

\usepackage{iopams}  
\makeatother

\usepackage{babel}

\begin{document}

\title{Temperature as an external field for colloid-polymer mixtures : ``quenching''
by heating and ``melting'' by cooling}

\author{Shelley L. Taylor}

\address{School of Chemistry, University of Bristol, Bristol, BS8 1TS, UK.}

\author{Robert Evans}

\address{H.H. Wills Physics Laboratory, University of Bristol, Bristol, BS8
1TL, UK.}

\author{C. Patrick Royall}

\address{School of Chemistry, University of Bristol, Bristol, BS8 1TS, UK.}

\ead{paddy.royall@bristol.ac.uk}

\address{Received April 20th, 2012}
\begin{abstract}
We investigate the response to temperature of a well-known colloid-polymer mixture.
At room temperature, the critical value of the second virial coefficient of the effective
interaction for the Asakura-Oosawa model predicts the onset of gelation with remarkable accuracy. Upon cooling the system, the effective attractions between colloids induced by polymer depletion are reduced, because the polymer radius of gyration is decreases as the
 $\theta$-temperature is approached. Paradoxically, this raises the effective temperature, leading to ``melting'' of colloidal gels. We find the Asakura-Oosawa model of effective colloid interactions with a simple description of the polymer temperature response provides a quantitative description of the fluid-gel transition. Further we present evidence for enhancement of crystallisation rates near the metastable critical point.
\end{abstract}

\section{Introduction}

Colloid-polymer mixtures occupy a special place in soft matter physics
\cite{poon2002, lekkerkerker}. The introduction of non-adsorbing polymer introduces an effective attraction between the colloids whose strength and range can be tuned by altering the concentration and molecular weight, respectively, of the polymer
\cite{lekkerkerker,asakura1954,asakura1958,vincent1972,vrij1976}. The existence of this entropy driven ‘depletion’ attraction
opens up a vast swathe of behaviour inaccessible to colloidal systems with purely
repulsive interactions, such that colloid-polymer mixtures may be
regarded as true ``model atomic systems'' \cite{poon2002}. The best known examples include liquid-gas phase separation \cite{poon2002, lekkerkerker} but there are also phenomena not seen in atomic systems such as gelation \cite{poon2002} and re-entrant glassy dynamics at high density \cite{pham2002}. Real-space analysis at the particle level has enabled direct observation of crystallisation \cite{dehoog2001}, and behaviour related to liquid-gas phase separation such as capillary wave fluctuations at (colloidal) liquid-gas interfaces \cite{aarts2004} and fluid critical phenomena \cite{royall2007c}. Moreover colloid-polymer mixtures with a relatively short-ranged attractive interaction can (crudely) model proteins and may exhibit two-step crystal nucleation phenomena \cite{savage2009}, to which we return below.

Theoretical treatments of colloid-polymer mixtures are based largely on the Asakura-Oosawa-Vrij (AO) model \cite{asakura1954,asakura1958,vrij1976} which treats the colloid-colloid interaction as that of hard-spheres (HS) and the polymer-polymer interaction as ideal, i.e. the polymer coils are assumed to be perfectly interpenetrating spheres. However, the polymer spheres have an excluded volume (hard) interaction with the HS colloids. This model binary (AO) mixture provides the simplest, zeroth-order description of the real mixture. It is characterized by the size ratio $q=\sigma_p/\sigma$, where sigma is the colloid diameter and $\sigma_p$ is the polymer sphere diameter. The colloid-polymer interaction is infinite for separations $r<(\sigma+\sigma_p)/2$. From simulation and theoretical studies it is well-known that for sufficiently large size ratios, $q \gtrsim 0.3$, the AO model exhibits phase separation into a colloid-rich (liquid) and a colloid-poor (gas) phase at sufficiently high polymer volume fractions. For smaller size ratios this phase transition becomes metastable w.r.t. the fluid-crystal transition\cite{gast1983,lekkerkerker1992,dijkstra1999}. The same trend in phase behaviour (with $\sigma_p$ set equal to twice the radius of gyration of the non-adsorbing polymer) is found in experimental studies\cite{ilett1995,poon2002,lekkerkerker}. In addition to predicting purely entropy driven fluid-fluid phase separation the AO model exhibits the elegant feature that for size ratios $q<(2/\sqrt 3-1)=0.1547$ the degrees of freedom of the ideal polymer can be integrated out exactly and the binary mixture maps formally to a one-component system of colloids described by an effective Hamiltonian containing only one and two-body (pair) contributions \cite{gast1983,dijkstra1999}. The former contribution plays no role in determining phase equilibrium or structure, for a uniform (bulk) fluid, but does determine the total pressure and compressibility \cite{dijkstra2000}.
The pair interaction between the HS colloids, given by integrating out the ideal polymer, is the standard AO potential:

\begin{equation} \beta u_{AO}(r)=
\cases{
\infty & for $r<\sigma$\\
- \phi_p  \frac{(1+q)^3}{q^3}  \\
\times \left[ 1-\frac{3r}{2(1+q)\sigma} +\frac{r^{3}}{2(1+q)^{3}\sigma^{3}}  \right]  & for $\sigma< r < \sigma+\sigma_p$ \\
0 & for $r  < \sigma+\sigma_p$ \\
} 
\label{eqAO}
\end{equation} 

\noindent $\beta$ is $1/k_{B}T$ where $k_{B}$ is Boltzmann's
constant and $T$ is temperature. Since the analysis is performed in the semi-grand ensemble the polymer volume fraction in the reservoir $ \phi_p = \pi \sigma_p^3 z_p / 6 $  appears in Eq. (vref{eqAO}).The polymer fugacity $z_{p}$ is equal to the number density $\rho_p$  of ideal polymers in the reservoir at the given chemical potential $\mu_p$. As noted already, in relating the AO model to experiment, one usually sets $\sigma_p = 2 R_G$ where $R_G$  is the polymer radius of gyration. Hitherto, most work on colloidal-polymer mixtures was carried out at a fixed
temperature, typically around 25 $^{\circ}$C. \emph{Effective}
temperature is varied by changing the interaction strength. The effective
temperature is inversely proportional to the depth of the attractive
well of the interaction potential in  (Eq. \ref{eqAO}), and is therefore fixed
for a given polymer reservoir density. Scanning a phase diagram then requires preparation of a considerable number of different samples. Conversely, in molecular
systems, interactions are usually constant over the (broad) temperature
range of interest, one sample is prepared and temperature is used as a control parameter.

In our present study we note that colloid-polymer mixtures can respond to temperature in an intriguing and counter-intuitive manner. The effective temperature in Eq. (\ref{eqAO}) is set by $z_{p}$ which in turn is equal to
the polymer number density for ideal polymers. Real systems approximate this behaviour very well \cite{royall2007jcp}. Thus the primary response of the system
to temperature is given by the response of the polymer depletant, since the 
polymer-polymer interactions are weak and colloid-colloid
interactions are athermal (hard sphere). Now, close to (but above) their theta
temperature $T^\theta$, polymers expand ($R_G$ increases)  upon heating (Fig. \ref{figRgT}) . This expansion has two effects : firstly, the polymer-colloid
size ratio $q$ increases, thereby increasing the range of attraction, and secondly, the polymer reservoir volume fraction  $\phi_p=\pi \rho_p \sigma_p^3 / 6$ also increases. Since the well-depth $- \beta u_{AO}(\sigma)= \pi \rho_p \sigma^3 q^2 (1+\frac{2}{3}q)/4$
 this means the effective temperature falls strongly for a modest increase in polymer size which leads to a paradoxical result, namely raising the temperature of a colloid-polymer mixture near $T^\theta$ brings about a strong \emph{effective cooling}. Although this effect has been
exploited to drive phase transitions in mixtures of colloidal rods
and polymers \cite{alsayed2004}, these temperature-dependent depletion
interactions have received relatively little attention. This is in
contrast to other means of controlling the attractive interactions between
colloids in-situ, such as the critical Casimir effect\cite{hertlein2008,guo2008,bonn2009}
and multiaxial electric fields \cite{elsner2009}. We note that in-situ control of attractive interactions, combined with particle-resolved studies, has the power to provide much new insight into a variety of phenomena, including phase transitions \cite{book}.

Here we make a quantitative experimental investigation, at the single-particle level, of the effect of temperature on an already well-studied colloid-polymer mixture. The elucidation of our results requires theoretical underpinning and we shall base this on the AO model described above. Specifically we investigate a mixture where the size ratio is about 0.2 but varies by 10 percent or so on changing temperature. The size ratio is such that the fluid-fluid transition is metastable w.r.t. the fluid-solid transition and therefore we consider out of equilibrium phenomena associated with metastable states.
A recent simulation study \cite{fortini2008} provides a helpful framework for placing our results in context. These authors study the effective one-component AO model, where the pair-potential is given by Eq(1), for $q=0.15$ which is in the regime where the mapping to the effective one-component description using only a pair potential  is exact. By changing the polymer reservoir density , equivalent to changing the depth of the attractive potential well, they determine both equilibrium and out-of equilibrium ‘phase diagrams’. More specifically, using Monte Carlo and Brownian dynamics, they investigate crystal nucleation and 
the onset of gelation in the vicinity of the metastable fluid-fluid binodal. They present convincing evidence that crystallization is enhanced by the binodal. We tackle the same issues in our experiments on a real colloid-polymer  mixture, seeking to ascertain what role proximity to the binodal plays in forming gels and in determining 
crystal nucleation rates.

It is well-known that in experiments
equilibrium is often not reached and in particular
gelation can occur. This phenomenon has been linked to spinodal decomposition
associated with colloidal gas-liquid condensation \cite{verhaegh1997,lu2008};
gels are supposed to form within the metastable fluid-fluid spinodal. It follows
that knowledge of the critical point is important in predicting where gelation might occur 
\cite{lu2008}.


The connection between critical density fluctuations and crystal nucleation
rates in short-ranged attractive systems, where the gas-liquid critical
point is metastable with respect to crystallisation, has received considerable
attention since it was elucidated by Ten Wolde and Frenkel \cite{tenWolde1997}. Critical fluctuations
are expected to enhance the nucleation rate and may be responsible
for the strong temperature dependence of nucleation rates found in globular
proteins \cite{galkin2000,vekilov2010}. A two-step nucleation process is envisaged 
where nuclei preferentially form in
fluctuations of high density since the surface tension between the
nucleus and the surrounding fluid is smaller. The reduction
in free energy barrier to nucleation associated with such density fluctuations 
has been measured in a 2D depletion system \cite{savage2009}. Here we investigate crystallisation in the neighbourhood of the metastable critical point.

This paper is organised as follows. In section \ref{sectionMethods} we first introduce the
experimental system, and discuss the response of the polymer component
to temperature. We then discuss relating
experiment and theory in terms of the AO model. In section \ref{sectionResults} we present results for \emph{(i)} the room-temperature phase diagram, \emph{(ii)} crystallisation around the metastable critical point and \emph{(iii)} the response of the system to temperature.  We conclude in section \ref{sectionConclusions}.

\section{Methods}
\label{sectionMethods}

\subsection{Experimental}
\label{sectionExperimental}
\begin{figure}[H]
\includegraphics[width=90mm]{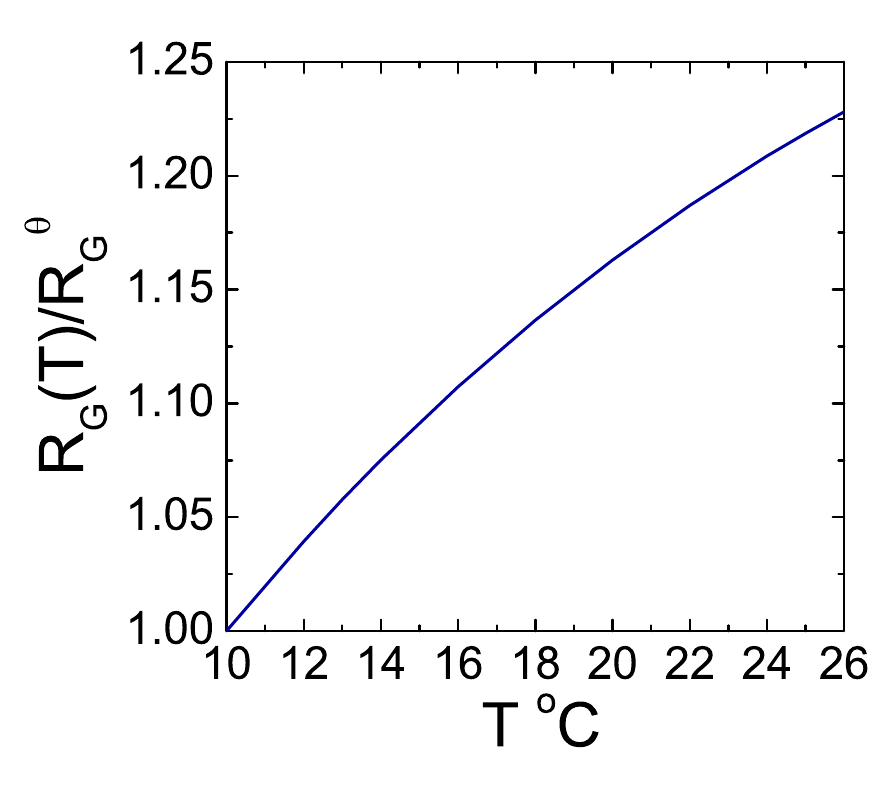} 
\caption{(color online) The radius of gyration $R_G$ of polystyrene. This is a fit,
Eq. (\ref{eqExpansion}), to experimental data \cite{berry1966}.
}
\label{figRgT} 
\end{figure}

Our experimental system is based on polymethyl methacrylate (PMMA) colloids.
The colloid diameter $\sigma=1080$ nm with polydispersity 4.6\%,
as determined from static light scattering. The polystyrene polymer
used has a molecular weight $M_{w}=8.5\times10^{6}$, which corresponds
to a radius of gyration of $R_{G}^{\theta}=95$ nm under $\theta$
conditions \cite{vincent1990}. This leads to a polymer-colloid size
ratio of $q(T=T^\theta)=0.176$. The colloids and polymer are dispersed in a solution
of \emph{cis} decalin, where we find $T^\theta$ 
of polystyrene is $10^{\circ}$C . We image this system at the
single particle level with a Leica SP5 confocal microscope. To this
microscope we have fitted a temperature stage which uses a Peltier chip
to cool from room temperature ($25{}^{\circ}$C) to the $\theta$-temperature
and below. Note that our experimental system is not density-matched. The gravitational
length $\lambda_{g}=k_{B}T/(mg)=1.96\sigma$, where $m$ is the buoyant
mass of the colloid and $g$ is the acceleration due to gravity. Sedimentation
therefore becomes an issue at long times, limiting our experimental
timescales to about one hour. For studies of crystallisation, we orient the sample capillaries perpendicular to gravity, mitigating its effect. We note that for this system a disordered layer forms on the capillary walls which inhibits heterogenous nucleation.

We estimate the effect of temperature on the radiius of gyration of the polymer as shown
in Fig. \ref{figRgT}. We use the following expression 

\begin{equation}
R_{G}(T)=R_{G}^{\theta}   \left[\sqrt{2}   \left(1-\exp \left( \frac{T^\theta-T}{\tau}   \right)   \right) +1 \right]
\label{eqExpansion}
\end{equation}

\noindent for $T\geq T^{\theta}$ which closely matches experimental
data over the relevant temperature range \cite{berry1966}. Here the parameter $\tau=20^\circ$ C. 


\subsection{Comparison with theory}
\label{sectionMapping}

As mentioned in the introduction we choose to interpret our experimental results within the framework of the simple AO model. A key ingredient is locating, in the AO phase diagram, the (metastable) fluid-fluid binodal for size- ratios $q\sim0.2$. There are computer simulation results for the binodal and its critical point for $q=0.1$ \cite{dijkstra1999, ashton2011} and for $q=0.15$ \cite{fortini2008}. Clearly in both cases $q<0.1547$ so the mapping to an effective one-component fluid that is described by only the pair potential Eq. (\ref{eqAO}) is exact. Although our experimental systems have $q\sim0.2$, within the context of AO we can expect three-body contributions to the effective Hamilitonian to play only a very small role. In order to ascertain the phase behaviour of the AO model it is tempting to turn to the free-volume theory \cite{lekkerkerker1992,dijkstra1999} which yields  simple recipes for calculating both fluid-solid and fluid-fluid phase equilibria for the binary AO mixture. This approximation is fairly successful for size- ratios $\gtrsim 0.4$. However, for $q=0.1$ free-volume theory provides a reasonably accurate description of fluid-solid coexistence \cite{dijkstra1999} but is quantitatively poor at describing the metastable fluid-fluid coexistence. Specifically it predicts a critical value of $\phi_p$ that is in reasonable agreement with simulation but a critical value of $\phi_c\sim 0.57$ that is unphysically large \cite{dijkstra1999}.

\begin{figure}[H]
 \includegraphics[width=140mm]{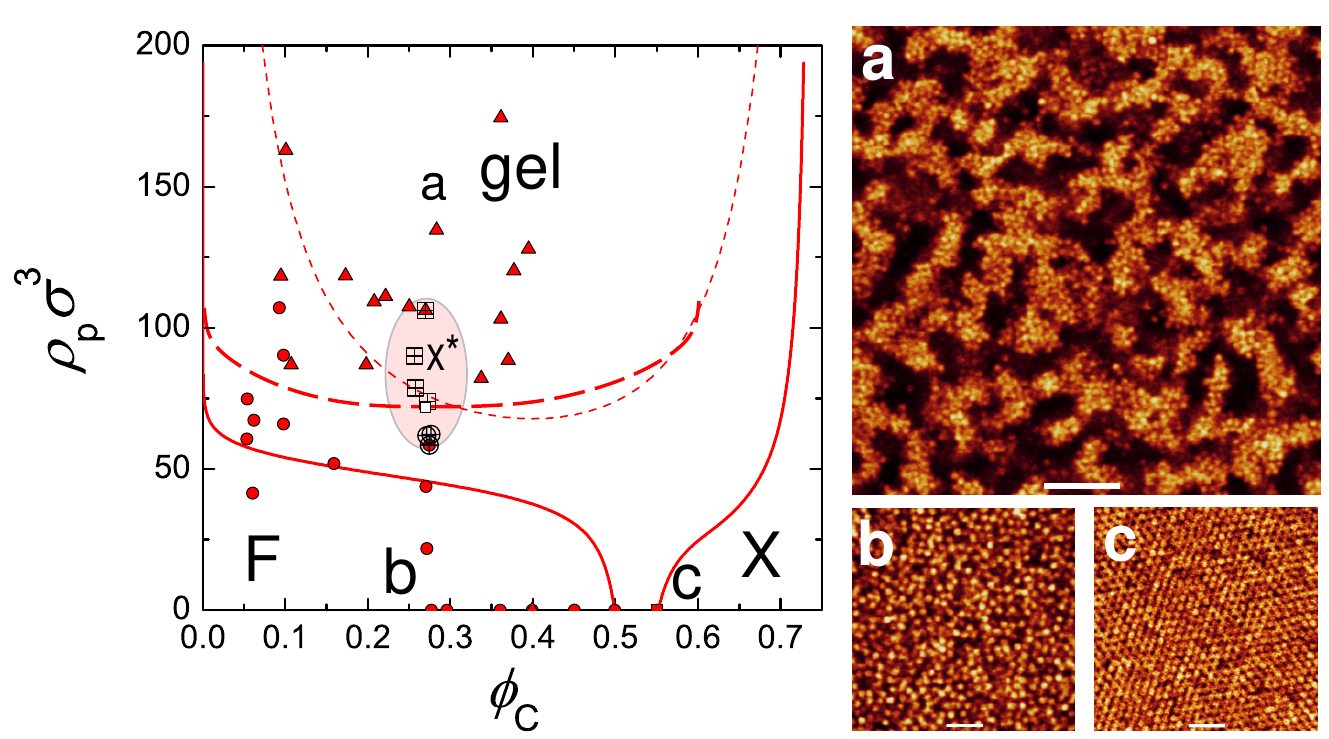} 
\caption{(color online) Phase diagram at room temperature in the colloid volume
fraction $ \phi_c $
and reservoir polymer number density $\phi_p$ plane. Symbols are experimental
data; circles are fluid (F), triangles are gels and squares are crystals (X).  Shaded area marked x$^*$ denotes samples which crystallised on the experimental timescale: hatched squares were intially gels, and hatched circles were initially fluids. The data are compared with theoretical predictions for the AO model with $q=0.214$ from free volume
theory for fluid-solid coexistence (solid lines) \cite{lekkerkerker1992}
and for the liquid-gas spinodal (short-dashed line) (Eq. \ref{schmidtSpin}). The unfilled square is the AO critical point according to our $B_2^{*AHS}$ criterion and the long dashed line
is a sketch of the accompanying spinodal. (a)-(c) are confocal
microscopy images of a gel (a), fluid (b) and (hard sphere) crystal
(c) at states in the phase diagram shown in the main panel. Bars=20 $\mu m$.
}
\label{figPdRoom} 
\end{figure}

In Fig. \ref{figPdRoom} we plot the fluid-fluid spinodal for $q=0.214$, the value that corresponds to the experimental system at room temperature $T=25^\circ$ $C$, calculated from free-volume theory using the analytical expression derived by Schmidt \emph{et al.} \cite{schmidt2002}:

\begin{equation}
\phi_{P}=\frac{\theta_{1}^{4}\theta_{2}/\phi_{c}}{\alpha\left(12\theta_{1}^{3}+15q\theta_{1}^{2}\theta_{2}+6q^2\theta_{1}\theta_{2}^{2}+q^3\theta_{2}^{3}\right)}
\label{schmidtSpin}
\end{equation}

\noindent where $\theta_{1}=(1-\phi_{c})$ and $\theta_{2}=(1+2\phi_{c})$
and $\alpha$ is the free volume fraction \cite{lekkerkerker1992,schmidt2002}.

One finds that the critical point is at about $\phi_c =0.40, \phi_p=0.35$. Once again the critical colloid fraction appears rather high. We shall argue that a more accurate value is $\phi_c \sim 0.27$.

Clearly a more reliable prescription is required to estimate the critical point and therefore the location of the binodal. For models like the present, where attractive interactions are short-ranged (sticky), Vliegenthart and Lekkerkerker \cite{vliegenthart2000} and Noro and Frenkel \cite{noro2000} argued that a useful estimate of the critical temperature (or interaction strength) could be obtained by considering the reduced, with respect to HS, second virial coefficient given by

\begin{equation}
B_{2}^{*}=\frac{3}{\sigma^3}\intop_{0}^{\infty}drr^{2}\left[1-
\exp\left(-\beta u(r)\right)\right]
\label{eqB2}
\end{equation}

\noindent 
where $u(r)$ is the pair potential. These authors proposed that for a wide class of model fluids $B_2^* \sim -1.5$ at the critical temperature. Later Largo and Wilding \cite{largo2006}
carried out simulations for effective (depletion) potentials calculated for additive binary HS systems with size-ratios $q=0.1$ and $0.05$. For these small ratios the effective pair potentials are similar to the AO potential [Eq. \ref{eqAO}] but with an additional repulsive barrier. 

For all the potentials they considered, Largo and Wilding found that the value of $B_2^*$ at criticality obtained from simulation was very close to $B_2^{*AHS} =-1.207$, i.e. the value reported for the adhesive hard sphere (AHS) model at its critical point \cite{miller2003}. Ashton applied the same criterion for the AO potential [Eq. (\ref{eqAO})] with $q=0.1$ \cite{ashton2011}. He found that his simulation result for the critical reservoir fraction $\phi_p$ was $0.249$, very close to the value $0.248$ given by the $B_2^{*AHS}$ criterion. We also considered the simulation results of Fortini et.al. \cite{fortini2008} for the AO model with $q=0.15$. In this case the critical value of $\phi_p$ is $\sim 0.316$ which is again close to the value $0.313$ from the $B_2^{*AHS}$ criterion. Since the size-ratios we consider are not vastly larger than those considered above, we chose to estimate the critical value of $\phi_p$ by calculating $B_2^*$ for the AO potential [Eq. (\ref{eqAO})] and employing the $B_2^{*AHS}$ criterion. In order to estimate the critical colloid fraction we used the mapping to the square-well potential proposed by Noro and Frenkel \cite{noro2000} to obtain an effective range. For the three temperatures, i.e. the three $q$ values, that we consider the estimate of the critical colloid fraction is about $0.27$ which happens to be equal to the AHS value \cite{miller2003}.

It is important to note that the simulations of the AO model, and of other models with short-ranged attraction, report broad, in $\phi_c$, gas-liquid coexistence curves extending to large values of $\phi_c$  and it is clear that extracting an accurate value for the critical colloid fraction can be difficult. Our resulting estimates of the AO model critical points are shown as open squares in Fig. \ref{figPdRoom} and Fig. \ref{figPdTemper} where we also sketch putative spinodals. Since our estimates are based on empirical recipes we remark that using the slightly higher value of $-1.174$ for the critical value of $B_2^{*AHS}$ ,reported by Largo et al. \cite{largo2008}, makes no discernible difference in our plots. Note also that their revised value of the critical packing fraction for AHS is $0.29$, only slightly bigger than the earlier result for AHS. Moreover were we to employ the original estimate for the critical value of $B_2^*$, namely $-1.5$, we find this results in only minor changes (about 4\%) to the critical value of $\phi_p$, for the size-ratios relevant for our systems.

\section{Results}
\label{sectionResults}

\subsection{Room-temperature behaviour}
\label{sectionRoom}

We begin our presentation of results with the phase diagram at ambient
conditions as shown in Fig. \ref{figPdRoom}. Throughout we work in
the colloid volume fraction $\phi_{c}$ and reservoir polymer number
density $\rho_{p}$ plane. We calculate $\rho_{p}$ following
the free-volume prescription for the AO model \cite{lekkerkerker1992} 
: $\rho_{p}=\rho_{p}^{exp}/\alpha$, where $\rho_{p}^{exp}$ is the experimental value for the polymer number density and $\alpha$ is the free-volume fraction entering Eq. (\ref{schmidtSpin}). In the free-volume approximation, $\alpha$ depends on $q$ and $\phi_c$ only \cite{dijkstra1999}. For a size-ratio $q \sim 0.2$ and colloid volume fractions up to $\phi_c \sim 0.4$ we expect this approximation to be accurate. 
Our choice of representation 
is motivated by two considerations : firstly, polymer number density
is conserved during heating and cooling (while polymer ``volume
fraction'' emphatically is not), and secondly, the reservoir representation
permits easy visual comparison with theoretical spinodal lines and critical
points.

We calculate theoretical phase boundaries as outlined in Section \ref{sectionMapping}: fluid-solid
coexistence is determined using free volume theory \cite{lekkerkerker1992} and
the metastable
fluid-fluid critical point is estimated using the $B_2^{*AHS}$ criterion.
Experimentally, along the hard sphere line $\phi_p=0$ ($x$-axis
in Fig. \ref{figPdRoom}) we find hard sphere crystallisation for
$\phi_{c}>0.54$. However, upon addition
of polymer, we found fluid states around the AO fluid-solid phase boundary and gelation
at higher polymer concentration. Only around the estimated AO critical point was
a pocket of states found which crystallised on the experimental timescale
of 6 days. This is the shaded region in Fig. \ref{figPdRoom}.

We already remarked that the fluid-gelation boundary has been identified with the fluid-fluid spinodal  \cite{verhaegh1997,lu2008}. In Fig. \ref{figPdRoom} we plot the spinodal calculated from free-volume theory , i.e. Eq. (\ref{schmidtSpin}), and a sketch of where the spinodal should be located for the AO model, based on our $B_2^{*AHS}$ criterion for the critical point. As mentioned earlier, free-volume theory grossly overestimates the colloid critical fraction for this size-ratio $q =0.214$ thus in making comparison with experiment it is appropriate to focus on the $B_2^{*AHS}$ result. We observe that the states identified as gels (triangles) all lie within the (putative) AO spinodal. Below the estimated AO fluid-fluid critical point we find only fluid states. We may conclude that the AO critical point provides an excellent indicator of the location of the experimental transition between fluid and gel states at this temperature.

\subsection{Critical enhancement of crystallisation}
\label{sectionXtal}

\begin{figure}[H]
\includegraphics[width=140mm]{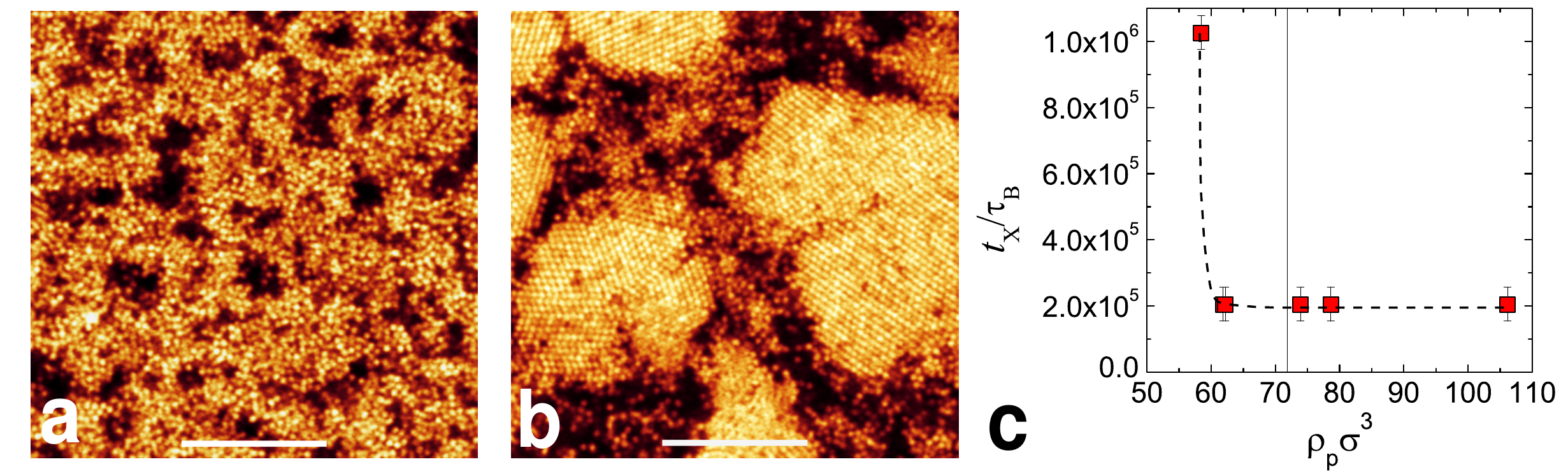} 
\caption{(color online) Crystallisation in colloid-polymer mixtures at room temperature. (a) immediately after preparation, (b) after 1 day. Bars=$20$ $\mu m$. (c) Crystallisation times in units of the Brownian time $\tau_B$ as a function of
polymer number density . Vertical line is the critical polymer number density estimated for the AO model, with $q=0.214$, from the $B_2^{*AHS}$ criterion. Dashed line is a guide to the eye.} 
\label{figXtal} 
\end{figure}

Crystallisation of colloid-polymer mixures is described in Fig. \ref{figXtal}. The crystals formed in the shaded region of Fig. \ref{figPdRoom} are markedly different from those formed in the absence of polymer in that the fluid is at a much lower colloid packing fraction. Crystallisation times vary from 1 to 6 days. We observe crystallisation only in a small ``pocket'' around the critical polymer number density $\rho_p^c \sigma^3 \sim 72$. 

Some samples which crystallised were fluids prior to freezing (hatched circles in Fig. \ref{figPdRoom}), while some were gels (hatched squares in Fig. \ref{figPdRoom}). 
That crystallisation is found to occur in the neighbourhood of the (metastable) critical point predicted by our $B_2^{*AHS}$ criterion is remarkable and we explore this aspect further in Fig. \ref{figXtal}(c) where we plot the crystallisation time as a function of polymer concentration. The crystallisation time $\tau_X$ is the time at which more than 50 \% of the sample had crystallised. As seen in Fig. \ref{figXtal}(b), crystallisation is rather clear. The unit of time employed is the Brownian time, defined as $\tau_B=\pi \eta \sigma^3 / k_B T$, where $\eta$ is the viscosity and is equal to $0.42$ $s$ for this system.
We find that in the (metastable) one-phase fluid region, moving further from criticality, the crystallisation time increases rapidly and takes values outside the experimental time-window. This is consistent with the two step nucleation scenario of ten Wolde and Frenkel 
\cite{tenWolde1997}.

\subsection{Response to temperature}
\label{sectionTemper}

\begin{figure}[H]
\includegraphics[width=140mm]{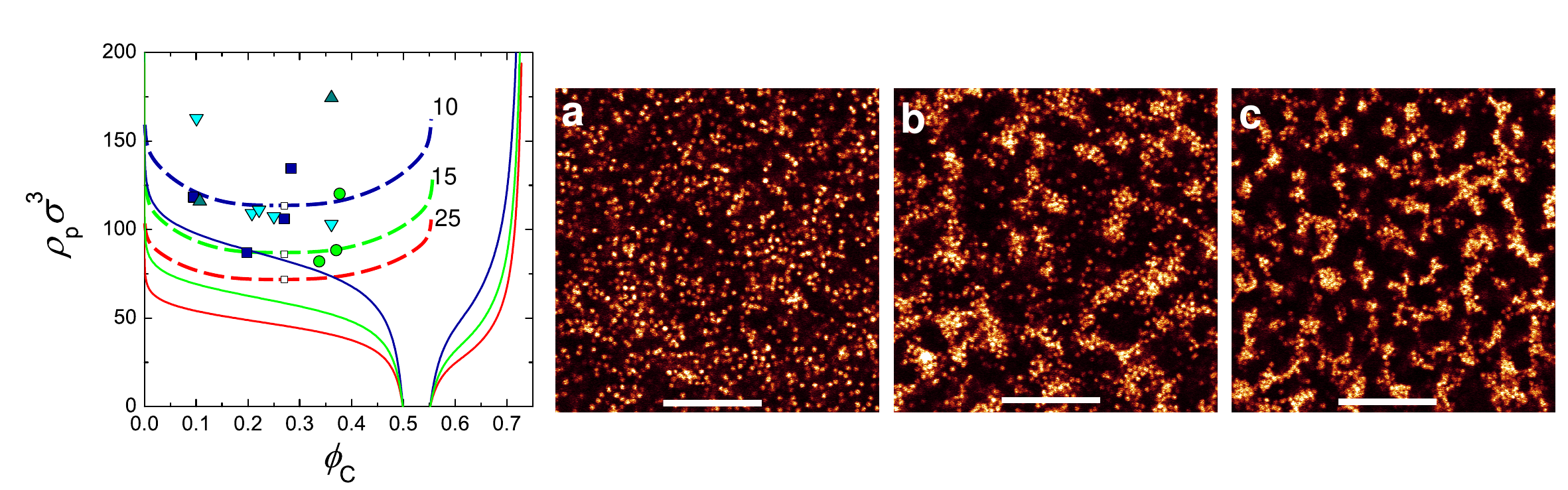} 
\caption{(color online) The fluid-gel transition at different temperatures. 
(a-c) shows a typical experiment where the transition temperature is found by
``quenching'' a  colloid-polymer mixture by heating. Here $\phi_c=0.107$ and $\rho_p=116 \sigma^{-3}$. (a) $8.5^\circ$C, fluid, (b) 
$10.2^\circ$C condensing, (c) $11.8^\circ$C, gel. The transition is then identified as around 
$10^\circ$C. 
In the main panel, experimental data indicate fluid-gel transitions at 
$10^\circ$C (blue squares), 
$11^\circ$C (turquoise up triangles),
$12^\circ$C (cyan down triangles) and
$16^\circ$C (green circles).
As in Fig. \ref{figPdRoom},  AO
free volume theory fluid-crystal coexistence lines are solid, 
and critical points (unfilled squares) follow our $B_2^{*AHS}$ criterion
with accompanying sketched spinodal (long dashed lines). These are shown for room temperature ($25^\circ$C),
$15^{\circ}$C and $10^{\circ}$C as indicated, corresponding to $q=0.214$, $0.197$ and $0.176$, respectively. Bars=20 $\mu m$.}
\label{figPdTemper}
\end{figure}

We now consider the effect of temperature on our system. The results are given in Fig. \ref{figPdTemper}. Using the temperature stage we cool the system to around $10^{\circ}$C. The images in Figs. \ref{figPdTemper}(a-c) show the effect of then gently heating the system. A metastable fluid (a) condenses (b) and finally forms a gel (c). 
The phase diagrams shown in the main panel pertain to the AO model and are obtained using the same prescriptions as in Fig. \ref{figPdRoom}. It is assumed that the only effect of temperature $T$ is to change the radius of gyration according to Eq. \ref{eqExpansion} and we calculate the size ratio using $q=2R_G(T)/\sigma$. We find $q= 0.214$, $0.197$ and $0.176$ for $T= 25$ $^\circ$C, $15$ $^\circ$C and $10$ $^\circ$C, respectively. Within the context of the free volume approximation the fluid-solid coexistence lines in the $\phi_c-\phi_p$ plane change little over this range of $q$ and we fixed these to be the lines for $q=0.18$. It is the scaling with $(\sigma/\sigma_p)^3$, from the polymer volume fraction to the polymer reservoir density, that gives rise to the variation shown in the figure. Although the experimental data show considerable scatter, which we attribute predominantly to sedimentation effects, there is reasonable overall agreement with the theoretical predictions. Specifically we find that a transition from a fluid to a gel as illustrated in Fig. \ref{figPdTemper} (a-c) occurs at temperatures broadly consistent with the location of the spinodals as predicted by our $B_2^{*AHS}$ criterion.
 It appears that the assumption of ideal polymer behaviour is a reasonable first step to treating the temperature response of colloid-polymer mixtures.

\section{Conclusions}
\label{sectionConclusions}

We have examined the room-temperature behaviour of a colloid-polymer mixture and its response to temperature quenches. To the best of our knowledge, these are the first particle-resolved studies of the latter. At ambient conditions, the fluid-gel transition is well described by
an estimate of the spinodal based on a second virial coefficient criterion for the effective one-component Asakura-Oosawa model. Although free-volume theory provides a reasonable description of fluid-crystal coexistence, we emphasize that this approximation predicts a spinodal which lies at unphysically large colloid volume fractions.

The response to temperature is consistent with our assumptions that the polymers can be  treated as ideal, and that the radius of gyration follows a simple fit to experimental results \cite{berry1966} [Eq. (\ref{eqExpansion})]. This opens the way to 3D particle-resolved studies of a variety of phenomena related to systems with attractive interactions which are tuneable \emph{in-situ}.

We find crystallisation on observable timescales close to the metastable critical point predicted by our $B_2^{*AHS}$ criterion. Assuming the mapping we have carried out is accurate, we find crystallisation in the metastable one-phase region, on a timescale which increases further from criticality, in addition to the metastable two-phase region. This contrasts with results from Brownian dynamics simulations where crystallisation was found only in the metastable gas-liquid two-phase region \cite{fortini2008,royall2012}. We attribute this to the very mch longer timescale accessed in the experiments.

Concerning the lack of crystallisation in the fluid-crystal coexistence regions, a question arises in the apparent discrepancy between our results
and Ilett \emph{et al.} \cite{ilett1995}, who found a closer agreement
with the prediction of the fluid-crystal boundary from free volume theory in a comparable
system (with size ratio $q=0.08$). However in their case, the colloid
diameter was $\sigma\approx400$ nm, compared to $\sigma\approx1080$
nm here . This has drastic consequences for the dynamics of the system,
as the time for a colloid to freely diffuse over its own diameter scales
with the cube of the particle size. Thus the effective timescales are around
20 times less in this work. Typical crystallisation times were around
6 hours in their case. This corresponds to 120 hours for our systems, which
is far beyond the experimental limits imposed by sedimentation of
around one hour. Observations such as this underline the challenges
for self-assembly in this size range and serve to emphasise the very
 dependence of timescales in these systems upon colloid size.

\textbf{Acknowledgments} SLT and CPR acknowledge the Royal
Society for financial support and EPSRC grant code EP/H022333/1 for
provision of the confocal microscope used in this work. Gregory N. Smith is kindly acknowledged for preliminary experiments. It is a pleasure to thank Daan Frenkel and Richard Sear for many stimulating discussions.

\section*{References}

\bibliographystyle{unsrt}

\end{document}